\documentclass[journal]{IEEEtran}
\usepackage{cite,graphicx,amsmath,psfrag,bm}
\usepackage[usenames,dvipsnames]{xcolor}
\usepackage{mathabx}
\usepackage{amsbsy}
\usepackage{subfigure}
\usepackage{epstopdf}
\usepackage{setspace}
\usepackage{tikz}
\usetikzlibrary{matrix}
\usepackage{pstool}
\graphicspath{{../}}

\begin{document}

\title{Generalized Sheet Transition Condition \\ FDTD Simulation of Metasurface}


\author{\IEEEauthorblockN{Yousef Vahabzadeh, Nima Chamanara, and Christophe Caloz,~\IEEEmembership{Fellow,~IEEE}}

\thanks{Yousef Vahabzadeh, Nima Chamanara, and Christophe Caloz are with the Department
of Electrical Engineering, Polytechnique de Montr$\acute{\mathrm{e}}$al, Montr$\acute{\mathrm{e}}$al,
QC, H3T 1J4 Canada (e-mail: yousef.vahabzadeh@polymtl.ca).}
\thanks{Manuscript received MONTH XX, 2017; revised MONTH XX, 2017.}}

\markboth{IEEE Transactions on Antennas and Propagation,~Vol.~X, No.~Y, Month~Z}%
{Shell \MakeLowercase{\textit{et al.}}: Simulation of Zero-Thickness Electromagnetic Sheet Using Finite difference Technique}

\maketitle

\begin{abstract}
We propose an FDTD scheme based on Generalized Sheet Transition Conditions (GSTCs) for the simulation of polychromatic, nonlinear and space-time varying metasurfaces. This scheme consists in placing the metasurface at virtual nodal plane introduced between regular nodes of the staggered Yee grid and inserting fields determined by GSTCs in this plane in the standard FDTD algorithm. The resulting update equations are an elegant generalization of the standard FDTD equations. Indeed, in the limiting case of a null surface susceptibility ($\chi_\text{surf}=0$), they reduce to the latter, while in the next limiting case of a time-invariant metasurface $[\chi_\text{surf}\neq\chi_\text{surf}(t)]$, they split in two terms, one corresponding to the standard equations for a one-cell ($\Delta x$) thick slab with volume susceptibility ($\chi$), corresponding to a diluted approximation ($\chi=\chi_\text{surf}/(2\Delta x)$) of the zero-thickness target metasurface, and the other transforming this slab in a real (zero-thickness) metasurface. The proposed scheme is fully numerical and very easy to implement. Although it is explicitly derived for a monoisotropic metasurface, it may be straightforwardly extended to the bianisotropic case. Except for some particular case, it is not applicable to dispersive metasurfaces, for which an efficient Auxiliary Different Equation (ADE) extension of the scheme is currently being developed by the authors. The scheme is validated and illustrated by five representative examples.
\end{abstract}

\IEEEpeerreviewmaketitle

\section{Introduction}

Metasurfaces are two-dimensional subwavelengthly thin bianisotropic structures consisting of scattering particles distributed on a substrate. They can perform a great diversity of electromagnetic wave transformations such as, for instance, broad-band focusing~\cite{pors2013broadband}, hologram generation~\cite{MTShologram}, anomalous refraction and reflection~\cite{Flat_Optics_CAPASO}, orbital angular momentum production~\cite{karimi2014generating}, space-surface wave manipulations~\cite{Achouri_arXiv_2016} and remote processing~\cite{Achouri_TAP_2015}.

\begin{figure}
\centering
\includegraphics[width=1\columnwidth]{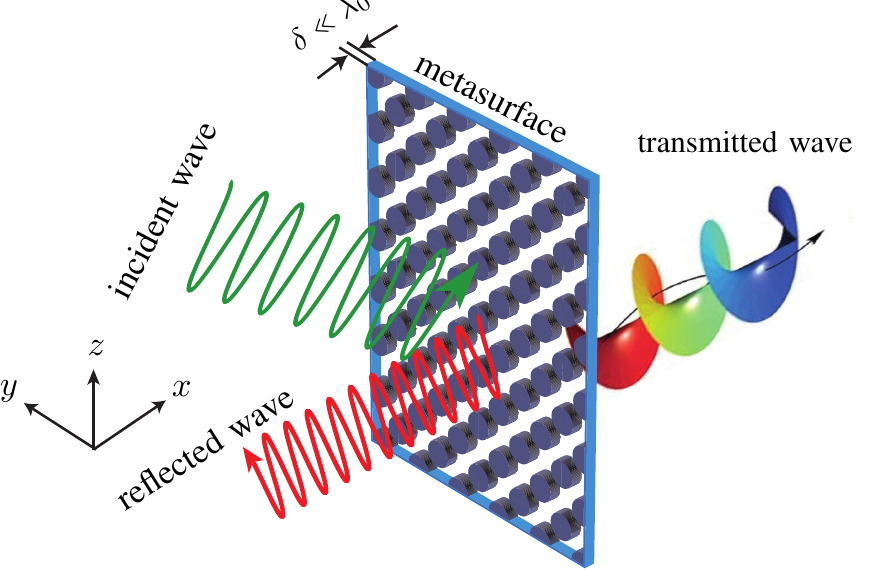}
  \caption{General representation of a metasurface, transforming an incident wave into a reflected wave and a transmitted wave. The metasurface thickness is much smaller than the operating wavelength, $\delta\ll\lambda_0$, and it may therefore be modeled as a zero-thickness sheet discontinuity.}
  \label{Fig:MSfig}
\end{figure}

Figure~\ref{Fig:MSfig} represents a general metasurface, transforming an incident wave into reflected and transmitted waves. As indicated in the figure, a metasurface is typically much thinner than the operation wavelength, i.e. $\delta\ll\lambda$. Therefore, it can be modeled as a zero-thickness sheet, which may simultaneously represent an electric and magnetic discontinuity. The metasurface may generally exhibit bianisotropy and variation in both space and time~\cite{MTSST}. The only known approach for exactly modeling a general metasurface is the Generalized Sheet Transition Condition (GSTC) technique~\cite{Idemen,kuester,MTSpolarizer,Grbic,Achouri_TAP_2015}.
Unfortunately, there is currently no commercial software including GSTCs and hence capable to simulate general metasurfaces. A metasurface may be approximated by an electrically thin slab with a volume susceptibility equal to the metasurface susceptibility divided by the thickness of the slab~\cite{Achouri_Hobart_2015,GSTCFDFD}. However, this technique is often inaccurate because a slab is a two-interface structure that cannot reduce to a single sheet structure. For example, as shown in~\cite{Achouri_Hobart_2015,GSTCFDFD}, the thin-slab approximation erroneously restricts the absorption level of a perfectly absorbing metasurface to $87\%$.

In~\cite{GRFDTD,2DEG}, a Finite-Difference Time-Domain (FDTD) based technique is proposed for the simulation of single magnetic field and single electric field sheet discontinuities corresponding for instance to graphene and 2D electron gas problems, respectively. In~\cite{GSTCFDFD} and~\cite{FEMKumar}, GSTC-based Finite difference Frequency Domain (FDFD) and Finite Element Method (FEM) schemes are introduced to simulate monochromatic, linear and time-invariant general-discontinuity (electric and magnetic) metasurfaces, respectively. While these approaches generally require meshing the three dimensional space, a GSTC-based spectral-domain integral-equation technique is proposed in~\cite{chamanara2017efficient} to reduce meshing to the surface of the metasurface. An equivalent RLC circuit based FDTD approach is presented in~\cite{FDFDShulbah1, FDFDShulbah2} to simulate space-time varying metasurfaces; however, this approach is limited to Lorentzian dispersion and inapplicable to problems involving extra scatterers.

In order to overcome the aforementioned limitations, and hence simulate general polychromatic, nonlinear and space-time varying metasurfaces placed in arbitrary scattering environments, we introduce here a completely general GSTC-FDTD scheme, that is easy to implement and that may be readily integrated in commercial softwares. This scheme is also shown to provides a fundamental generalization of the conventional FDTD technique.

The paper is organized as follows. Section~\ref{sec:synthesis} briefly describes the metasurface synthesis equations in the time domain and the frequency domain. Upon this basis, Sec.~\ref{sec:analysis} derives the proposed GSTC-FDTD scheme for 1D and 2D problems. Benchmark and illustrative examples are provided in Sec.~\ref{sec:examples}. Finally, conclusions are given in Sec.~\ref{sec:conclusions}.

\section{Metasurface Synthesis Equations}\label{sec:synthesis}
The metasurface is assumed to be positioned in the $y-z$ plane of a cartesian coordinate system as shown in Fig.~\ref{Fig:MSfig}. It is illuminated by an arbitrary wave from the left side and transforms this incident wave into arbitrary reflected and transmitted waves at the left and right sides of the structure, respectively. The metasurface is best modelled by electric and magnetic surface susceptibility tensors in the frequency domain or the time domain.

\subsection{Frequency Domain}\label{subsec:freqdom}
Without loss of generality and for simplicity, we assume that the metasurface does not support any normal electric and magnetic polarization currents, so that the corresponding polarization densities are zero, i.e. $P_x=0$ and $M_x=0$. In this case, the frequency-domain (time harmonic dependency $e^{+j\omega t}$) bianisotropic GSTCs read~\cite{karim}
\begin{subequations}\label{GSTC}
  \begin{align}\label{GSTC1}
  \left(
  \begin{array}{c}
    -\Delta \tilde{H}_z \\
    \Delta \tilde{H}_y \\
  \end{array}
\right) &=j\omega\varepsilon_0 \left(
                        \begin{array}{cc}
                          \tilde{\chi} _{\textrm{ee}}^{yy} & \tilde{\chi} _{\textrm{ee}}^{yz} \\
                          \tilde{\chi} _{\textrm{ee}}^{zy} & \tilde{\chi} _{\textrm{ee}}^{zz} \\
                        \end{array}
                      \right)
                      \left(
                        \begin{array}{c}
                          \tilde{E}_{y,\textrm{av}} \\
                          \tilde{E}_{z,\textrm{av}} \\
                        \end{array}
                      \right)\\\notag
                      &+j\omega\sqrt{\varepsilon_0 \mu_0} \left(
                                                           \begin{array}{cc}
                                                             \tilde{\chi} _{\textrm{em}}^{yy} & \tilde{\chi} _{\textrm{em}}^{yz} \\
                                                             \tilde{\chi} _{\textrm{em}}^{zy} & \tilde{\chi} _{\textrm{em}}^{zz} \\
                                                           \end{array}
                                                         \right)
                                                         \left(
                                                           \begin{array}{c}
                                                             \tilde{H}_{y,\textrm{av}} \\
                                                             \tilde{H}_{z,\textrm{av}} \\
                                                           \end{array}
                                                         \right),
  \end{align}
  \begin{align}\label{GSTC2}
  \left(
  \begin{array}{c}
    -\Delta \tilde{E}_y \\
    \Delta \tilde{E}_z \\
  \end{array}
\right) &=j\omega\mu_0 \left(
                        \begin{array}{cc}
                          \tilde{\chi} _{\textrm{mm}}^{zz} & \tilde{\chi} _{\textrm{mm}}^{zy} \\
                          \tilde{\chi} _{\textrm{mm}}^{yz} & \tilde{\chi} _{\textrm{mm}}^{yy} \\
                        \end{array}
                      \right)
                      \left(
                        \begin{array}{c}
                          \tilde{H}_{z,\textrm{av}} \\
                          \tilde{H}_{y,\textrm{av}} \\
                        \end{array}
                      \right)\\\notag
                      &+j\omega\sqrt{\varepsilon_0 \mu_0} \left(
                                                           \begin{array}{cc}
                                                             \tilde{\chi}_{\textrm{me}}^{zz} & \tilde{\chi} _{\textrm{me}}^{zy} \\
                                                             \tilde{\chi}_{\textrm{me}}^{yz} & \tilde{\chi} _{\textrm{me}}^{yy} \\
                                                           \end{array}
                                                         \right)
                                                         \left(
                                                           \begin{array}{c}
                                                             \tilde{E}_{z,\textrm{av}} \\
                                                             \tilde{E}_{y,\textrm{av}} \\
                                                           \end{array}
                                                         \right),
\end{align}
\end{subequations}
where $\Delta\tilde{\psi}_u=\tilde{\psi}_u^\textrm{tr}-(\tilde{\psi}_u^\textrm{ref}+\tilde{\psi}_u^\textrm{inc})$ and $\tilde{\psi}_{u,\textrm{av}}=\frac{(\tilde{\psi}_u^\textrm{inc}+\tilde{\psi}_u^\textrm{ref})+\tilde{\psi}_u^\textrm{tr}}{2}$ with $u=x, y$ or $z$ and $\tilde{\psi}$ representing the spectral electric or magnetic fields, and the superscripts "inc", "tr" and "ref" denote the incident, transmitted and reflected waves, respectively, and the $\tilde{\chi}$'s represent the generally frequency-dependent surface susceptibility tensor terms.

The synthesis procedure consists in determining the susceptibility tensors in~\eqref{GSTC} for specified sets (possibly multiple ones for multiple transformations) of fields $(\psi_u^\textrm{inc},\psi_u^\textrm{ref},\psi_u^\textrm{tr})$~\cite{karim}. Since Eqs.~\eqref{GSTC} represent a linear system of equations, this generally (always in the case of a full-rank system) leads to explicit solutions, \mbox{$\bar{\bar{\chi}}_\textrm{ab}=\bar{\bar{\chi}}_\textrm{ab}(y,z)$} ({\small a,b$=$e,m}). Moreover, if the fields are specified in closed-form, then the solution expressions are also closed form.

\subsection{Time Domain}\label{sec:time_dom}

The time-domain GSTCs are obtain by taking the inverse Fourier transform of~\eqref{GSTC}. Assuming that the metasurface is non-dispersive, i.e. that the $\tilde{\chi}$'s in~\eqref{GSTC} do not depend on~$\omega$, the terms $\tilde{\chi}\tilde{\psi}_u(\omega)$ simply transform to $\chi\psi_u(t)$ where $\chi=\tilde{\chi}$, without involving any convolution. The time-domain counterpart of~\eqref{GSTC} is then obtained by just replacing $j\omega$ with the operator $\frac{d}{dt}$. For notational simplicity and without loss of generality, we write here only the monoisotropic version of the resulting equations:
\begin{subequations}\label{TGSTC}
  \begin{align}\label{TGSTC1}
    \Delta H_y&=\varepsilon_0\frac{d\left[\chi_\textrm{ee}^{zz}E_{z,\textrm{av}} \right]}{dt},\\\label{TGSTC2}
    \Delta E_z&=\mu_0\frac{d\left[\chi_\textrm{mm}^{yy}H_{y,\textrm{av}} \right]}{dt}.
  \end{align}
\end{subequations}

Note that $\frac{d}{dt}$ generally operates on the \emph{products} of the susceptibility tensors and average fields. Such products, when both terms are time-dependent, lead to the generation of new frequencies. Here, we are generally looking for solutions of the form \mbox{$\bar{\bar{\chi}}_\textrm{ab}(y,z;t)$}. In contrast to the frequency-domain case, these solutions are generally neither explicit nor closed-form. So, the synthesis problem is substantially more complicated in the time domain than in the frequency domain.

\section{GSTC-FDTD Scheme}\label{sec:analysis}

We shall restrict to the case of monoisotropic metasurfaces. We first address the 1D problem, where the metasurface reduces to a point, and is hence a 0D metasurface or ``metapoint'', and next extend the scheme to the corresponding 2D problem, where the metasurface reduces to a segment, and is hence a 1D metasurface or ``metasegment''. Upon this foundation, the extension to the bianisotropic and 3D cases is straightforward although tedious.

\subsection{1D Analysis}\label{sec:analysis_1D}

\begin{figure}
\centering
\includegraphics[width=1\columnwidth]{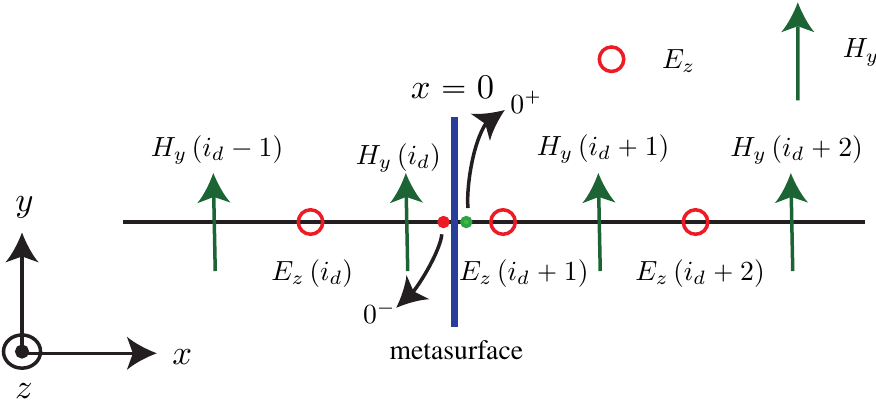}
  \caption{Computational scheme for the 1D analysis of a 0D metasurface. The metasurface, characterized by surface susceptibility tensors, is placed at a \emph{virtual} node between sampling points of the 1D staggered Yee FDTD grid. The filled red and green circles represent electric and magnetic virtual nodes just before and just after the metasurface, i.e. at $x=0^-$ and $x=0^+$, respectively. The incident wave propagates in the $x-$direction.}
  \label{Fig:1DFDFD}
\end{figure}

We consider here the 1D computational problem of a 0D metasurface, represented in Fig.~\ref{Fig:1DFDFD}. This still represent the practical 3D problem of a uniform 2D metasurface perpendicularly illuminated by a TEM plane wave. The metasurface is positioned in the 1D FDTD staggered Yee grid as shown in Fig.~\ref{Fig:1DFDFD}, i.e. at $x=0$, between the two neighboring cells $i_d$ and $i_d+1$, where $i_d$ is the number of the cell before the metasurface. The incident wave illuminates the metasurface from the left of the metasurface ($x<0$) while the right side ($x>0$) is the transmitted-wave region.

Everywhere except at the nodes around the discontinuity, one can use the conventional FDTD update equations. These equations read~\cite{FDTD_Susan}
 \begin{subequations}\label{RegularFDTD}
   \begin{align}\label{RegularFDTD1}
     &H_y^{n+\frac{1}{2}}\left(i\right)=H_y^{n-\frac{1}{2}}\left(i\right)+\frac{\Delta t}{\mu_0\Delta x}\left[E_z^n\left(i+1\right)-E_z^n\left(i\right)\right],\\\label{RegularFDTD2}
     &E_z^n \left( i\right)=E_z^{n-1}\left( i\right)+\frac{\Delta t}{\varepsilon_0\Delta x}\left[H_y^{n-\frac{1}{2}}\left(i \right) -H_y^{n-\frac{1}{2}}\left( i-1\right)\right]
   \end{align}
 \end{subequations}
where $n$ represents the discrete time, $t=n\Delta t$, with $\Delta t$ being the time step.

In contrast, special treatment must be applied to the nodes around the discontinuity. Consider the update equation for $E_z^n\left( i_d+1\right)$ in Fig.~\ref{Fig:1DFDFD}, that would conventionally depend on the surrounding fields $H_y^{n-1/2}\left( i_d\right)$ and $H_y^{n-1/2}\left( i_d+1\right)$. Because of the presence of the discontinuity between cells $i_d$ and $i_d+1$, $H_y^{n-1/2}\left( i_d\right)$ \emph{cannot} be readily used as such in~\eqref{RegularFDTD2} for otherwise nothing would be accounting for the effect of the discontinuity. To remedy this situation, we introduce a \emph{virtual magnetic node} (H-node) at the right side of the metasurface and rewrite the update equation involving this new node as
\begin{align}\label{Eyeqstep1}
  E_z^n \left( i_d+1\right)=&E_z^{n-1}\left( i_d+1\right)+\\\notag
  &\frac{\Delta t}{\varepsilon_0\Delta x} \left[ H_y^{n-\frac{1}{2}}\left(i_d+1 \right) -H_y^{n-\frac{1}{2}}\left( 0^+\right) \right],
\end{align}
where $H_y^{n-\frac{1}{2}}\left( 0^+\right)$ is unknown. To determine this quantity, we invoke the GSTC equation~\eqref{TGSTC1}, whose time-domain discretization reads
\begin{equation}\label{Eyeqstep2}
  H_y^{n-\frac{1}{2}}\left( 0^+\right)=H_y^{n-\frac{1}{2}}\left( i_d\right)+\frac{\varepsilon_0}{\Delta t}[\left(\chi_\textrm{ee}^{zz}E_{z,\textrm{av}} \right)^n-\left( \chi_\textrm{ee}^{zz}E_{z,\textrm{av}} \right)^{n-1}],
\end{equation}
where the effect of the metasurface is now accounted for in the second term of the right-hand side of this equation. Substituting~\eqref{Eyeqstep2} in~\eqref{Eyeqstep1}, and grouping identical components, yields

\begin{subequations}
\label{Eyeqstep3}
\begin{align}\label{Eyeqstep3a}
  E_z^n \left( i_d+1\right)&A_\textrm{ee}^{zz,n}=E_z^{n-1}\left( i_d+1\right)A_\textrm{ee}^{zz,n-1}+\\\notag
  &\frac{\Delta t}{\varepsilon_0\Delta x} \left[ H_y^{n-\frac{1}{2}}\left(i_d+1 \right) -H_y^{n-\frac{1}{2}}\left(i_d\right) \right]-\\\notag
  &\frac{\chi_\textrm{ee}^{zz,n}}{2\Delta x}E_z^n\left(i_d \right)+\frac{\chi_\textrm{ee}^{zz,n-1}}{2\Delta x}E_z^{n-1}\left(i_d \right),
\end{align}
with
\begin{equation}
A_\textrm{ee}^{zz,n}=1+\frac{\chi_\textrm{ee}^{zz,n}}{2\Delta x}.
\end{equation}
\end{subequations}

Now consider the update equation for $H_y^{n+\frac{1}{2}}\left(i_d\right)$, that would conventionally depend on the surrounding fields $E_z^{n}\left(i_d\right)$ and $E_z^{n}\left(i_d+1\right)$. Following the same argument as for $E_z^n\left( i_d+1\right)$, we introduce the \emph{virtual electric node} (E-node) at the left side of the metasurface and rewrite the update equation~\eqref{RegularFDTD1} involving this new node as
 \begin{equation}\label{Hxeqstep1}
   H_y^{n+\frac{1}{2}}\left(i_d\right)=H_y^{n-\frac{1}{2}}\left(i_d\right)+\frac{\Delta t}{\mu_0\Delta x}\left[E_z^n\left(0^-\right)-E_z^n\left(i_d\right) \right],
 \end{equation}
next invoke the GSTC equation~\eqref{TGSTC2}, whose time-domain discretization reads
  \begin{align}\label{Hxeqstep2}
  E_z^n\left( 0^- \right)&=E_z^n\left(i_d+1 \right)-\\\notag  &\frac{\mu_0}{\Delta t}\left[ \left( \chi_\textrm{mm}^{yy}H_{y,\textrm{av}}\right)^{n+\frac{1}{2}} - \left(\chi_\textrm{mm}^{yy}H_{y,\textrm{av}}\right)^{n-\frac{1}{2}} \right],
  \end{align}
and finally substitute~\eqref{Hxeqstep2} into~\eqref{Hxeqstep1} to obtain
\begin{subequations}\label{Hxeqstep3}
   \begin{align}\label{Hxeqstep3a}
   H_y^{n+\frac{1}{2}}&\left(i_d\right)A_\textrm{mm}^{yy,n+\frac{1}{2}}=H_y^{n-\frac{1}{2}}\left(i_d\right)A_\textrm{mm}^{yy,n-\frac{1}{2}}+\\\notag
   &\frac{\Delta t}{\mu_0\Delta x}\left[E_z^n\left(i_d+1\right)-E_z^n\left(i_d\right) \right]-\\\notag &\frac{\chi_\textrm{mm}^{yy,n+\frac{1}{2}}}{2\Delta x}H_y^{n+\frac{1}{2}}\left( i_d+1\right)+ \frac{\chi_\textrm{mm}^{yy,n-\frac{1}{2}}}{2\Delta x}H_y^{n-\frac{1}{2}}\left( i_d+1\right).
 \end{align}
 with
\begin{equation}
A_\textrm{mm}^{yy,n}=1+\frac{\chi_\textrm{mm}^{yy,n}}{2\Delta x}.
\end{equation}
\end{subequations}

Equations~\eqref{Eyeqstep3} and~\eqref{Hxeqstep3} are the update equations for the nodes around the metasurface discontinuity. In the absence of metasurface, corresponding to $\chi_\textrm{ee}^{zz,n}=\chi_\textrm{mm}^{yy,n}=0$, these equations reduce to their conventional counterpart~\eqref{RegularFDTD}, as expected. A next limit case is that of a time-invariant metasurface, for which $\chi_\textrm{ee}^{zz,n}$ and $\chi_\textrm{mm}^{yy,n}$ are independent of time. In this case, Eq.~\eqref{Hxeqstep3}, for instance, reduces to
   \begin{align}\label{Nochi}
   H_y^{n+\frac{1}{2}}&\left(i_d\right)=H_y^{n-\frac{1}{2}}\left(i_d\right)+\\\notag
   &\frac{\Delta t}{\mu_0\mu_r\Delta x}\left[E_z^n\left(i_d+1\right)-E_z^n\left(i_d\right) \right],
 \end{align}
with $\mu_\textrm{r}=1+\frac{\chi_\textrm{mm}^{xx}}{2\Delta x}$. This equation corresponds to its conventional counterpart for a ``metasurface'' modeled by a thin slab of \emph{volume} permeability $\mu_r$, which may be understood as a dilution of the metasurface \emph{surface} permeability, $\mu_\textrm{r,\textrm{surf}}$, across two grid cells, with the overall thickness of $2\Delta x$, namely $\mu_\textrm{r}=\mu_\textrm{r,\textrm{surf}}/(2\Delta x)$. A similar time-invariant susceptibility result is naturally obtained for~\eqref{Eyeqstep3}. Thus, the last two terms in~\eqref{Eyeqstep3a} and~\eqref{Hxeqstep3a} may be interpreted as corrections of the corresponding conventional equations to model a zero-thickness sheet.

\subsection{2D Analysis}\label{sec:analysis_2D}
\begin{figure}
\includegraphics[width=1\columnwidth]{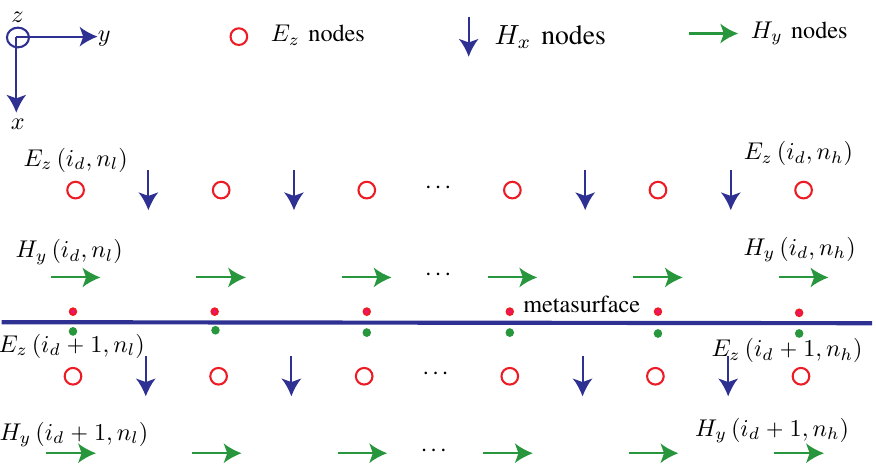}
\caption{Computational scheme for the 2D analysis of a 1D metasurface. The metasurface, characterized by surface susceptibility tensors, is placed at a \emph{virtual array} of nodes between sampling points of the 2D staggered Yee FDTD grid. The filled red and green circles represent electric and magnetic virtual nodes just before and just after the metasurface, i.e. at $x=0^-$ and $x=0^+$, respectively. The incident wave propagates in the $xy-$plane.}\label{fig:FDTDYEE}
\end{figure}

We now consider the 2D computational problem of a 1D metasurface, represented in Fig.~\ref{fig:FDTDYEE}. This still represents the practical 3D problem of a 1D-nonuniform (e.g. phase-gradient) 2D metasurface illuminated by an oblique TM$_z$ or TE$_z$ plane wave. Without loss of generality, we solve here the TM$_z$ problem, with $E_z, H_x$ and $H_y$ being the non-zero field components,  while a similar scheme straightforwardly applies to the TE$_z$ case. The metasurface is positioned in the 2D FDTD staggered Yee grid as shown in Fig.~\ref{fig:FDTDYEE}, where $i_d$ is the number of the cell before the metasurface and $j_d=n_l:n_h$ represents the metasurface extension in the $y-$direction. The incident wave illuminates the metasurface in the $xz-$ plane.

Consider the update equation for $E_z^n\left(i_d+1,j_d \right)$ at the nodes around the discontinuity which would conventionally require $H_y^{n-\frac{1}{2}}\left(i_d,j_d\right)$. However, as in the 1D case, this field quantity cannot readily be used as such, because that would miss the effect of the metasurface, and we therefore introduce a \emph{virtual array of magnetic nodes} at the lower side of the discontinuity. This leads to the update equation
\begin{align}\label{Ezeqstep1}
  E_z^n &\left( i_d+1,j_d\right)=E_z^{n-1}\left( i_d+1,j_d\right)+\\\notag
  &\frac{\Delta t}{\varepsilon_0\Delta x} \left[ H_y^{n-\frac{1}{2}}\left(i_d+1,j_d\right)-H_y^{n-\frac{1}{2}}\left( 0^+,j_d\right) \right]+\\\notag
  &\frac{\Delta t}{\varepsilon_0\Delta y}\left[H_x^{n-\frac{1}{2}}\left(i_d+1,j_d\right)-H_x^{n-\frac{1}{2}}\left( i_d+1,j_d\right) \right],
\end{align}
which is the 2D counterpart of~\eqref{Eyeqstep1}. In this expression, $H_y^{n-\frac{1}{2}}\left( 0^+,j_d\right)$ is determined from the GSTC equation~\eqref{TGSTC1} as
\begin{align}\label{Ezeqstep2}
  H_y^{n-\frac{1}{2}}\left( 0^+,j_d\right)=&H_y^{n-\frac{1}{2}}\left( i_d,j_d\right)+\\\notag &\frac{\varepsilon_0}{\Delta t}\left[\left(\chi_\textrm{ee}^{zz}E_{z,\textrm{av}} \right)^n-\left( \chi_\textrm{ee}^{zz}E_{z,\textrm{av}} \right)^{n-1}\right],
\end{align}
which is the 2D counterpart of~\eqref{Eyeqstep2}. Finally, substituting~\eqref{Ezeqstep2} into~\eqref{Ezeqstep1} and regrouping yields
\begin{subequations}\label{Ezeqstep3}
\begin{align}
  E_z^n &\left( i_d+1,j_d\right)A_\textrm{ee}^{zz,n}=E_z^{n-1}\left( i_d+1,j_d\right)A_\textrm{ee}^{zz,n-1}+\\\notag
  &\frac{\Delta t}{\varepsilon_0\Delta x} \left[ H_y^{n-\frac{1}{2}}\left(i_d+1,j_d \right) -H_y^{n-\frac{1}{2}}\left(i_d,j_d\right) \right]+\\\notag
  &\frac{\Delta t}{\varepsilon_0\Delta y}\left[H_x^{n-\frac{1}{2}}\left(i_d+1,j_d \right) -H_x^{n-\frac{1}{2}}\left(i_d+1,j_d\right) \right]-\\\notag
  &\frac{1}{2\Delta x}\left[ \chi_\textrm{ee}^{zz,n}E_z^n\left(i_d,j_d\right)  -   \chi_\textrm{ee}^{zz,n-1}E_z^{n-1}\left(i_d,j_d\right)   \right].
\end{align}
with
\begin{equation}
A_\textrm{ee}^{zz,n}=1+\frac{\chi_\textrm{ee}^{zz,n}}{2\Delta x},
\end{equation}
\end{subequations}
the 2D counterpart of~\eqref{Eyeqstep3}.

Similarly, one obtains for the 2D counterpart of~\eqref{Hxeqstep1}, \eqref{Hxeqstep2} and~\eqref{Hxeqstep3}, assuming a \emph{virtual array of electric nodes} at the upper side of the discontinuity,
\begin{equation}\label{Hyeqstep1}
\begin{split}
  &H_y^{n+\frac{1}{2}}\left(i_d,j_d\right)=H_y^{n-\frac{1}{2}}\left(i_d,j_d\right)+\\
  &\qquad\qquad\qquad\frac{\Delta t}{\mu_0\Delta x}\left[E_z^n\left(0^-,j_d\right)-E_z^n\left(i_d,j_d\right) \right],
\end{split}
\end{equation}
  \begin{align}\label{Hyeqstep2}
  E_z^n\left( 0^-,j_d \right)&=E_z^n\left( i_d+1,j_d \right)-\\\notag
  &\frac{\mu_0}{\Delta t}\left[ \left( \chi_\textrm{mm}^{yy}H_{y,\textrm{av}}\right)^{n+\frac{1}{2}} - \left(\chi_\textrm{mm}^{yy}H_{y,\textrm{av}}\right)^{n-\frac{1}{2}} \right],
  \end{align}
and
   \begin{align}\label{Hyeqstep3}
   H_y^{n+\frac{1}{2}}&\left(i_d,j_d\right)A_m^{y,n+\frac{1}{}}=H_y^{n-\frac{1}{2}}\left(i_d,j_d\right)A_m^{y,n-\frac{1}{2}}+\\\notag &\frac{\Delta t}{\mu_0\Delta x}\left[E_z^n\left(i_d+1,j_d\right)-E_z^n\left(i_d,j_d\right) \right]-\\\notag &\frac{\chi_\textrm{mm}^{yy,n+\frac{1}{2}}}{2\Delta x}H_y^{n+\frac{1}{2}}\left( i_d+1,j_d\right)+\\\notag &\frac{\chi_\textrm{mm}^{yy,n-\frac{1}{2}}}{2\Delta x}H_y^{n-\frac{1}{2}}\left( i_d+1,j_d\right).
 \end{align}

\section{Note on Dispersive Metasurfaces}\label{Sec:Disp_MTs}

Section~\ref{sec:analysis}, based on~\eqref{TGSTC}, assumed \emph{non-dispersive} time-varying metasurfaces, and the time dependency of the susceptibilities corresponding to the time variance is apparent from the $n$-dependency of the susceptibility terms $\chi_\textrm{ee}^{zz,n}$ in~\eqref{Eyeqstep3} and \eqref{Ezeqstep3} and $\chi_\textrm{mm}^{yy,n-1/2}$ in~\eqref{Hxeqstep3} and \eqref{Hyeqstep3}. However, metasurfaces are often dispersive, due to their typically resonant nature, and the corresponding time-varying susceptibilities would then be depending also on frequency, i.e. $\tilde{\chi}=\tilde{\chi}(\omega;t)$ in Sec.~\ref{subsec:freqdom}. In such a case, the virtual node concept introduced in this work for accommodating GSTCs in FDTD is still applicable, but must be accompanied with a standard FDTD treatment of dispersion, with update equations that are specific to the dispersion considered~\cite{Susan}. A complete extension of the present work to the case of dispersive metasurfaces will be the object of a next paper.

It turns out that the scheme of Sec.~\ref{sec:analysis} still applies to particular dispersion forms, specifically to those forms that allow proper staggered Yee grid discretization in time. Let us see this in the forthcoming examples.

First consider a metasurface absorbing part of a normally incident wave and transmitting the rest of it. In general, such a metasurface is represented by susceptibilities of the form $\tilde{\chi}_\textrm{ee}^{zz}(\omega)=\tilde{\chi}_\textrm{mm}^{yy}(\omega)=\frac{\kappa}{j\omega}$, where $\kappa$ only depends on the reflection and transmission coefficients\footnote{Assuming TM$_z$ operation ($\tilde{H}_z=\tilde{E}_x=\tilde{E}_y=0$) and monoisotropy ($\bar{\bar{\chi}}_\textrm{em}=\bar{\bar{\chi}}_\textrm{me}=0$, $\bar{\bar{\chi}}_\textrm{ee}=\hat{z}\tilde{\chi}_\textrm{ee}^{zz}\hat{z}$ and $\bar{\bar{\chi}}_\textrm{mm}=\hat{y}\tilde{\chi}_\textrm{mm}^{yy}\hat{y}$), the metasurface synthesis equations~\eqref{GSTC} reduce to $\Delta \tilde{H}_y=j\omega\varepsilon_0\tilde{\chi}_\textrm{ee}^{zz}\tilde{E}_{z,\textrm{av}}$ and \mbox{$\Delta \tilde{E}_z=j\omega\mu_0\tilde{\chi}_\textrm{mm}^{yy}\tilde{H}_{y,\textrm{av}}$}. Inserting the specified fields, $\tilde{E}_z^\textrm{inc}=e^{-jk_0x}$, $\tilde{E}_z^\textrm{ref}=\Gamma e^{+jk_0x}$ and $\tilde{E}_z^\textrm{tr}=Te^{-jk_0x}$, into these equations at \mbox{$x=0$} yields $\tilde{\chi}_\textrm{ee}^{zz}=\frac{1}{j\omega\varepsilon_0}\frac{\tilde{H}_y^\textrm{tr}-(\tilde{H}_y^\textrm{inc}+\tilde{H}_y^\textrm{ref})}{[\tilde{E}_z^\textrm{tr}+(\tilde{E}_z^\textrm{inc}+\tilde{E}_z^\textrm{ref})]/2}=\frac{2c_0}{j\omega}\frac{T-1-\Gamma}{T+1+\Gamma}$ and $\tilde{\chi}_\textrm{mm}^{yy}=\frac{1}{\omega\mu_0}\frac{\tilde{E}_z^\textrm{tr}-(\tilde{E}_z^\textrm{inc}+\tilde{E}_z^\textrm{ref})}{[\tilde{H}_y^\textrm{tr}+(\tilde{H}_y^\textrm{inc}+\tilde{H}_y^\textrm{ref})]/2}=\frac{2c_0}{j\omega}\frac{T-1-\Gamma}{T+1+\Gamma}$.\label{fn:Abs_MS_synth}}. Substituting these susceptibilities into~\eqref{GSTC} yields
\begin{align}\label{S4_Ex1_1}
  \Delta \tilde{H}_y&=\varepsilon_0\kappa \tilde{E}_{z,\textrm{av}},\\
  \Delta \tilde{E}_z&=\mu_0\kappa \tilde{H}_{y,\textrm{av}},
\end{align}
which represents the very particular case where the frequency dependency has vanished. The corresponding discretized time-domain equations read
\begin{subequations}\label{S4_Ex1_2}
  \begin{align}\label{S4_descr1}
    \Delta H_y^{n-\frac{1}{2}}&=\frac{\varepsilon_0\kappa}{2}\left(E_{z,\textrm{av}}^n+E_{z,\textrm{av}}^{n-1}\right),\\\label{S4_descr2}
    \Delta E_z^n&=\frac{\mu_0\kappa}{2}\left(H_{y,\textrm{av}}^{n+\frac{1}{2}}+H_{y,\textrm{av}}^{n-\frac{1}{2}}\right).
  \end{align}
\end{subequations}
From this point, one can follow exactly the same method as in Sec.~\ref{sec:analysis} with just replacing~\eqref{Eyeqstep2} and~\eqref{Hxeqstep2} for the 1D case and~\eqref{Ezeqstep2} and~\eqref{Hyeqstep2} for the 2D case by~\eqref{S4_descr1} and~\eqref{S4_descr2}, respectively.

Consider now the Lorentzian metasurface, $\tilde{\chi}_{ee}^{zz}(\omega)=\frac{1}{-\omega^2+j\gamma\omega+\omega_0^2}$. In this case, without any loss of generality and for the sake of simplicity, we use only $\tilde{\chi}_{ee}^{zz}$, the same result being applicable for the cases involving other components of the susceptibility tensors. Substituting this susceptibility into~\eqref{GSTC} yields
\begin{equation}\label{S4_Ex2_1}
  \Delta \tilde{H}_y=\frac{j\omega\varepsilon_0}{-\omega^2+j\gamma\omega+\omega_0^2} \tilde{E}_{z,\textrm{av}},
\end{equation}
and the discretized time-domain counterpart of this equation is obtained by replacing $j\omega$ with $\frac{d}{dt}$, which yields
\begin{equation}\label{S4_EX2_2}
\begin{split}
  &\frac{\Delta H_y^{n+\frac{1}{2}}-2\Delta H_y^{n-\frac{1}{2}}+\Delta H_y^{n-\frac{3}{2}}}{2\Delta t}+\gamma\frac{\Delta H_y^n -\Delta H_y^{n-1}}{\Delta t}+\\
  &\qquad\qquad\qquad\omega_0^2 \Delta H_y^{n-\frac{1}{2}}=\varepsilon_0\frac{E_{z,\textrm{av}}^n-E_{z,\textrm{av}}^{n-1}}{2}.
\end{split}
\end{equation}
This equation requires H-field samples at both half-integer and integer times whereas, according the staggered nature of the Yee grid, the H-field is available only at half-integer times. Equation~\eqref{S4_EX2_2} is thus incompatible with the staggered Yee grid and standard FDTD algorithm. Resolving this issue would require the introduction of auxiliary functions~\cite{Susan} and, as previously mentioned, this will be done in a next paper. Equation~\eqref{S4_EX2_2} would be directly compatible with the FDTD algorithm only in the case $\gamma=0$, which would correspond to a non-causal lossless and hence impractical dispersive system.

Generalizing from the above examples, we see the technique proposed in this paper is directly applicable only when the first and second derivatives of the magnetic or electric field do not simultaneously appear in the time-domain GSTC equation or, equivalently, when $j\omega$ and $(j\omega)^2$ do not simultaneously appear in the denominator or nominator of the frequency-domain susceptibility.

\section{Note on Nonlinear Metasurfaces}\label{Sec:Nonlin_MTs}

The proposed GSTC-FDTD technique is also applicable to nonlinear metasurfaces, where the susceptibility function depends on orders of the electric or magnetic fields larger than one and where the susceptibility tensors reach an order corresponding to the order of the nonlinearity. The nonlinear extension of the proposed technique is straightforward. It has in fact already been implemented in a work by some of the authors of this paper specifically dealing with nonlinear metasurfaces and the details are available in the appendix of~\cite{Karim2017Nonlinear}.

\section{Benchmark and Illustrative Examples}\label{sec:examples}

This section presents the following five application examples of the GSTC-FDTD scheme presented in Sec.~\ref{sec:analysis}:

\begin{enumerate}
  \item 1D analysis of constant-susceptiblity reflection-less 0D metasurface benchmarked with analytical result;
  \item 1D analysis of sinusoidally time-varying reflection-less 0D metasurface;
  \item 2D analysis of 1D metasurface representing a graphene sheet benchmarked with literature results;
  \item 2D analysis of 1D space-time reflection-less varying and dispersive metasurface benchmarked with synthesis specification.
  \item 1D analysis of reflection-less half-absorbing and half-transmitting 0D dispersive metasurface benchmarked with synthesis specification;
\end{enumerate}

\subsection{Example 1}

Consider the 1D computational problem of a 0D reflectionless metasurface with constant and equal susceptibilities $\chi_\textrm{ee}^{zz}=\chi_\textrm{mm}^{yy}=5$. It may be shown that in the case of equal monoisotropic susceptibilities, the metasurface is reflection-less~\cite{karim}. The following analytical closed-form transmitted field is found from~\eqref{TGSTC} for the (necessarily) normally incident and transverse electromagnetic (TEM) plane wave $H_y^\textrm{inc}=\sin\left( \omega t-k_0x\right)$:
\begin{align}\label{Exact1DEx1}
  H_y^\textrm{tr}=&\frac{1-\frac{k_0^2}{4}\chi_\textrm{mm}^{yy}\chi_\textrm{ee}^{zz}}{1+\frac{k_0^2}{4}\chi_\textrm{mm}^{yy}\chi_\textrm{ee}^{zz}} \sin\left( \omega t-k_0x\right)\\\notag &+\frac{1}{\eta_0}\frac{\epsilon_0\omega\chi_\textrm{ee}^{zz}}{1+\frac{k_0^2}{4}\chi_\textrm{ee}^{zz}\chi_\textrm{mm}^{yy}}\cos\left(\omega t-k_0x \right).
\end{align}
\begin{figure}
\includegraphics[width=1\columnwidth]{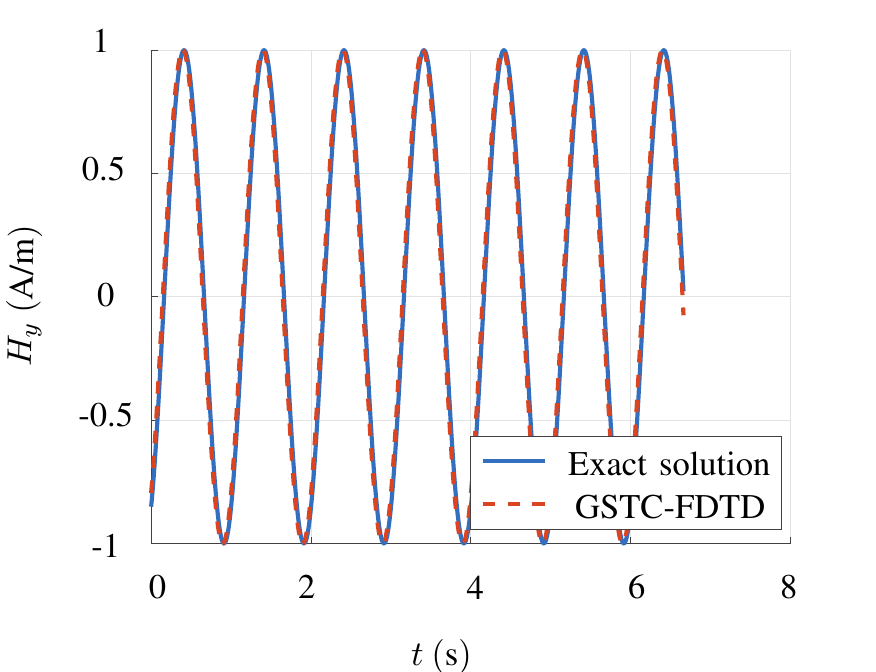}
\caption{Example 1: Comparison of the GSTC-FDTD and exact analytical [Eq.~\eqref{Exact1DEx1}]  transmitted waveforms. The solid red and dashed black line represent the exact and simulated results, respectively. The number of the nodes per wavelength is $N_\textrm{res}=30$.}\label{Fig:ConstX1d_Ex1}
\end{figure}
The GSTC-FDTD simulation and exact analytical results for this problem are compared in Fig.~\ref{Fig:ConstX1d_Ex1}. In the steady state, the two results feature undistinguishable amplitude waveforms in the scale range of the figure, with a zero reflected field, consistently with the design.

\subsection{Example 2}

The second example is the 1D computational problem of a 0D reflection-less time-varying metasurface with susceptibilities $\chi_\textrm{ee}^{zz}=\chi_\textrm{mm}^{yy}=\chi=1+\sin\left(\omega t\right)$, where the two susceptibilities have been made equal for zero reflection~\cite{karim}, with normally incident TEM wave \mbox{$E_z=\sin\left(\omega t\right)$}. Due to the time-varying nature of the metasurface, generation of new time-harmonic frequencies are expected, as may be understood from inspecting~\eqref{TGSTC}.
\begin{figure}
\includegraphics[width=1\columnwidth]{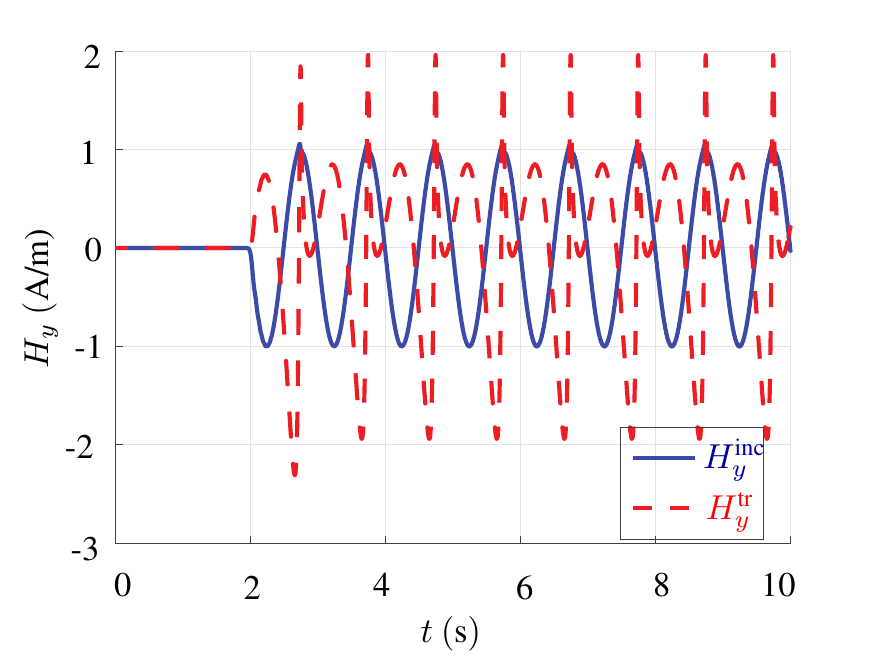}
\caption{Example 2: GSTC-FDTD transmitted (dashed red line) and incident (blue line) waves for the metasurface discontinuity with sinusoidal time-varying susceptibility profile.}\label{Fig:Ex3_Tvarying}
\end{figure}

The simulation result is plotted in Fig.~\ref{Fig:Ex3_Tvarying}. Reflection has been verified to be zero and the emergence of new frequencies may be deduced from the distorted shaped of the transmitted wave in the figure. Unfortunately, this example cannot be benchmarked due to the inexistence of appropriate commercial code.

\subsection{Example 3}

We shall now demonstrate the applicability of our GSTC-FDTD scheme to a graphene sheet~\cite{GRFDTD} 2D-analyzed as a 1D metasurface. Since the scheme applies to both electric and magnetic field discontinuities, it is expected to automatically work for the case of single magnetic-field discontinuity represented by graphene.

\begin{figure}
\centering
\includegraphics[width=1\columnwidth]{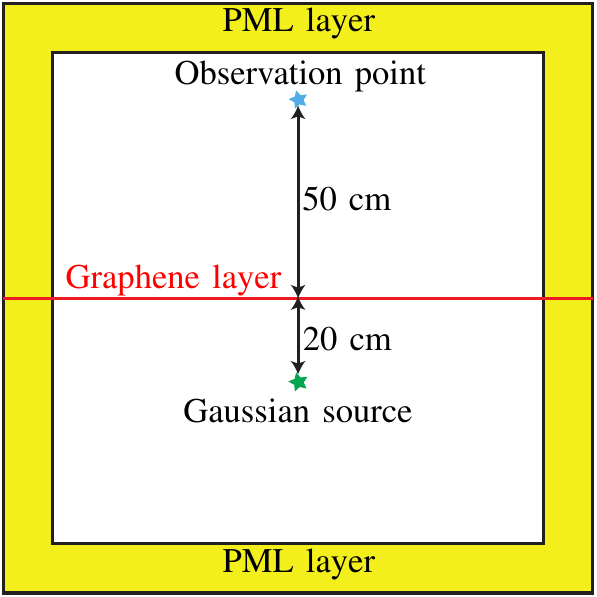}
  \caption{Example 3: Position of the source and observation point with respect to the graphene sheet.}
  \label{Fig:GrapheneEx2}
\end{figure}

Since graphene is continuous in terms of the electric field (while being discontinuous in terms of the magnetic field), it is naturally positioned at the $E_z\left(i_d+1,j_d\right)$-nodes in the FDTD grid of Fig.~\ref{fig:FDTDYEE}. We have thus to compute the magnetic field at the nodes before and after the graphene sheet. From~\cite{GRFDTD}, the discretized equation for (non-magnetized) graphene is
\begin{equation}\label{graphene1}
  E_z^n=C^+\left(H_y^{0^+} -H_y^{0^-}\right)^{n+\frac{1}{2}}+C^-\left(H_y^{0^+} -H_y^{0^-}\right)^{n-\frac{1}{2}},
\end{equation}
with $C^+=\frac{\Delta t+2\tau}{2\Delta t\sigma_0}$ and $C^-=\frac{\Delta t-2\tau}{2\Delta t\sigma_0}$.
Replacing $E_z^n\left( i_d+1,j_d\right)$ with~\eqref{graphene1} in~\eqref{Hyeqstep1}, and grouping identical components, yields

\begin{align}\label{graphene2}
 H_y^{n+\frac{1}{2}}&\left( i_d+1,j_d\right)D^+=H_y^{n-\frac{1}{2}}\left( i_d+1,j_d\right)D^-+\\\notag
 &\frac{\Delta t}{\mu_0\Delta x}\left[ E_z^n\left(i_d+2,j_d\right)+C^+H_y^{n+\frac{1}{2}}\left(i_d,j_d\right)+\right.\\\notag &\quad \left. C^-H_y^{n-\frac{1}{2}}\left(i_d,j_d\right)  \right],
\end{align}
with $D^+=1+\frac{\Delta t}{\mu_0\Delta x}C^+$ and $D^-=1-\frac{\Delta t}{\mu_0\Delta x}C^-$. Following the same procedure for $H_y^{n+\frac{1}{2}}\left( i_d,j_d\right)$ yields
\begin{align}\label{graphene3}
 H_y^{n+\frac{1}{2}}&\left( i_d,j_d\right)D^+=H_y^{n-\frac{1}{2}}\left( i_d,j_d\right)D^-+\\\notag
 &\frac{\Delta t}{\mu_0\Delta x}\left[ E_z^n\left(i_d,j_d\right)+C^-H_y^{n+\frac{1}{2}}\left(i_d+1,j_d\right)+\right.\\\notag &\quad \left. C^+H_y^{n-\frac{1}{2}}\left(i_d+1,j_d\right)  \right].
\end{align}
As in~\cite{GRFDTD}, we set the incident wave as the gaussian-derivative wave $I_z(t)=-\sqrt{2e}\beta\left(t-t_0\right)\exp\left(-\beta^2\left( t-t_0^2\right)^2 \right)\:\left(\mu \textrm{A}\right)$ with $\beta=10/t_0$, $t_0=1\:\textrm{ps}$ and $e$ the electron charge. The position of the source and observation point is shown in Fig.~\ref{Fig:GrapheneEx2}.

\begin{figure}
\centering
\includegraphics[width=1\columnwidth]{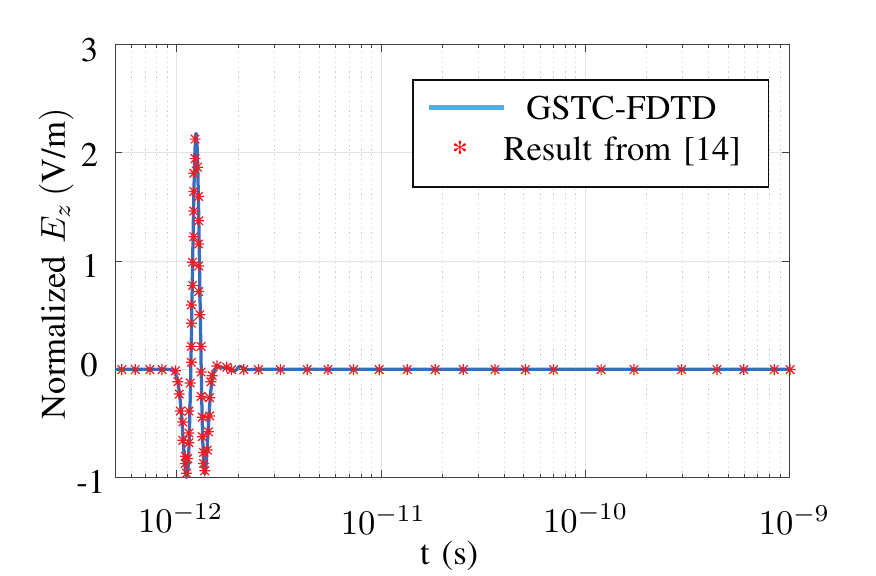}
  \caption{Example 3: Comparison of the GSTC-FDTD result with the result of~\cite{GRFDTD}. $N_\textrm{res}=50$.}
  \label{Fig:Graphene_result}
\end{figure}

Simulation results are plotted in Fig.~\ref{Fig:Graphene_result} and satisfactorily compared with the result of~\cite{GRFDTD}.
\subsection{Example 4}

\begin{figure}
\centering
\includegraphics[width=1\columnwidth]{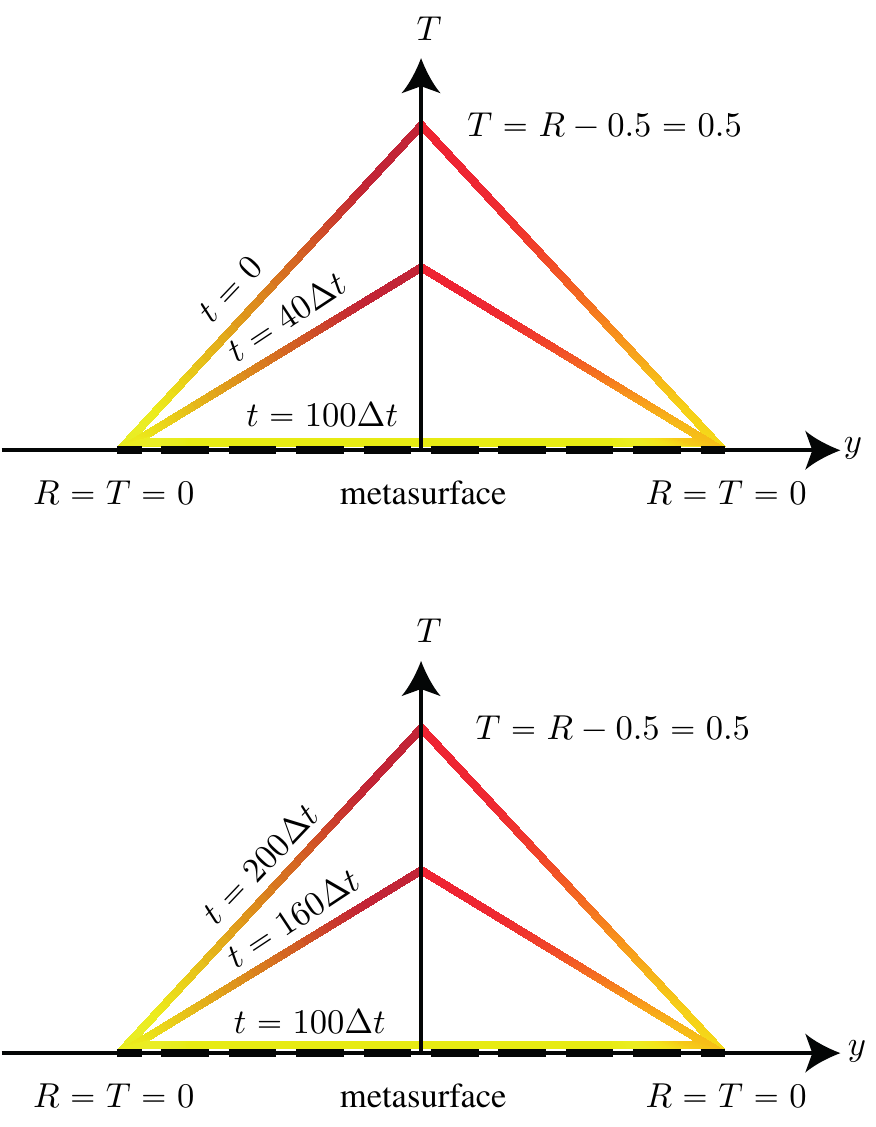}
  \caption{Example 4: Susceptibility variation of the metasurface in space and time. Initially, the metasurface is full absorbing at the edges and half-power transmitting at the middle. After 100 time steps, it is full absorber every where and then, at the end of the period, it recovers its initial state. Both time and space variations are linear.}
  \label{fig:st_profile}
\end{figure}

The next example is the 2D analysis of the 1D space-time reflection-less metasurface described in Fig.~\ref{fig:st_profile}. This metasurface is synthesized to linearly oscillate in space and time from fully absorbing to half transmitting as shown in the figure. From the metasurface synthesis equations~\eqref{GSTC} and Footnote~\ref{fn:Abs_MS_synth}, the susceptibilities for an absorbing and half transmitting metasurfaces are \mbox{$\tilde{\chi}_{mm}^{yy}=\tilde{\chi}_{ee}^{zz}=\frac{-2c_0}{j\omega}$} and \mbox{$\tilde{\chi}_{mm}^{yy}=\tilde{\chi}_{ee}^{zz}=\frac{-2c_0}{3j\omega}$}, respectively, or generally $-\frac{2c_0}{j\omega\alpha(t)}$, where we set the periodic time variation with triangular period
\[  \alpha(t)= \left\{
\begin{array}{ll}
      1+\frac{1}{50\Delta t}t, & \text{for  } 0\leq t \leq 100\Delta t, \vspace{2mm} \\
      3-\frac{1}{50\Delta t}(t-100\Delta t), & \text{for  } 100\Delta t \leq 200\Delta t,
\end{array}
\right. \]
with $\Delta t=15.349\:\textrm{ps}$. According to this relation, $\alpha(0)=1$, $\alpha(100\Delta t)=3$, $\alpha(200\Delta t)=1$, and so on, periodically. Therefore, the corresponding device is time-varying metasurface linearly oscillating between full power absorption ($\alpha=1$) and half power transmission ($\alpha=3$). The time-varying GSTC for the metasurface is thus
\begin{subequations}
   \begin{align}
     \Delta H_y=\frac{-2c_0\epsilon_0}{\alpha(t)}E_{z,\textrm{av}},\\
     \Delta E_z=\frac{-2c_0\mu_0}{\alpha(t)}H_{y,\textrm{av}}.
   \end{align}
 \end{subequations}

Following the procedure outlined in Sec.~\ref{sec:analysis_2D} and Sec.~\ref{Sec:Disp_MTs}, the update equation for $H_y^{n+\frac{1}{2}}\left( i_d,j_d\right)$ is found as
\begin{align}
   &H_y^{n+\frac{1}{2}}\left(i_d,j_d\right)F^+=H_y^{n-\frac{1}{2}}\left(i_d,j_d\right)F^-+\\\notag &\frac{\Delta t}{\mu_0\Delta x}\left[E_z^n\left(i_d+1,j_d\right)-E_z^n\left(i_d,j_d\right) \right]-\\\notag
   &\frac{\Delta tc_0}{2\Delta x\alpha(t)} \left[ H_y^{n+\frac{1}{2}}\left( i_d+1,j_d\right)+H_y^{n-\frac{1}{2}}\left( i_d+1,j_d\right)\right],
\end{align}
with $F^+=1+\frac{\Delta tc_0}{2\Delta x\alpha(t)}$ and $F^-=1-\frac{\Delta tc_0}{2\Delta x\alpha(t)}$, while that for $E_y^{n}\left(i_d+1,j_d \right)$ reads
\begin{align}
   &E_z^{n}\left(i_d+1,j_d\right)R^+=E_z^{n-1}\left(i_d+1,j_d\right)R^--\\\notag &\frac{\Delta t}{\epsilon_0\Delta x}\left[H_y^{n-\frac{1}{2}}\left(i_d+1,j_d\right)-H_y^{n-\frac{1}{2}}\left(i_d,j_d\right) \right]+\\\notag
   &\frac{\Delta t}{\epsilon_0\Delta y}\left[ H_x^{n-\frac{1}{2}}\left( i_d+1,j_d\right)-H_x^{n-\frac{1}{2}}\left( i_d+1,j_d-1\right)\right]-\\\notag
   &\frac{\Delta tc_0}{6\Delta z\alpha(t)} \left[ E_y^{n}\left( i_d,j_d\right)+E_y^{n+1}\left( i_d,j_d\right)\right],
\end{align}
with $R^{\pm}=1\pm\frac{c_0\Delta t}{2\Delta x\alpha(t)}$.

\begin{figure}[ht]
\subfigure[]{%
\includegraphics[width=1\columnwidth]{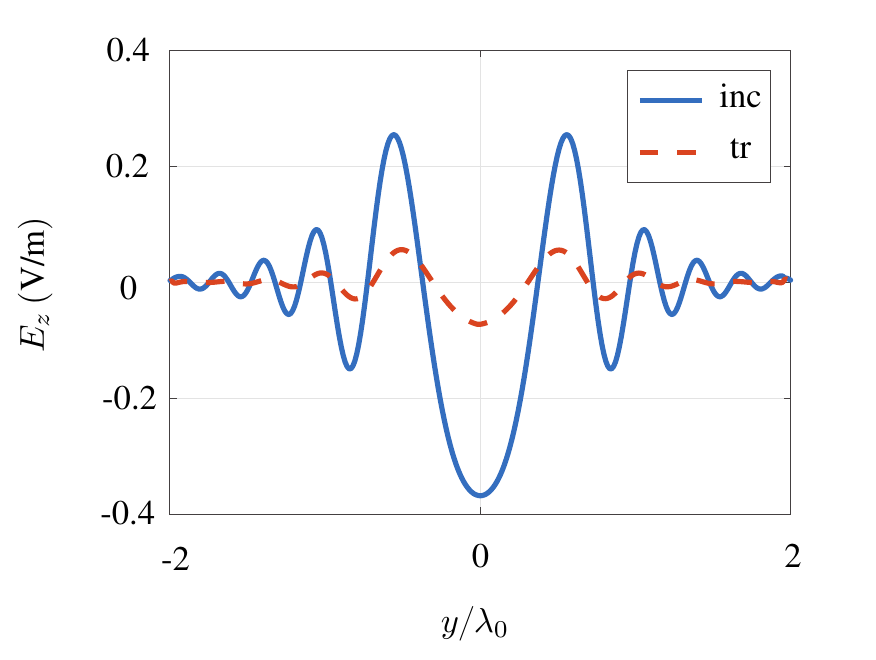}
   \label{Ex5_1_A}}%

   \subfigure[]{%
   \includegraphics[width=1\columnwidth]{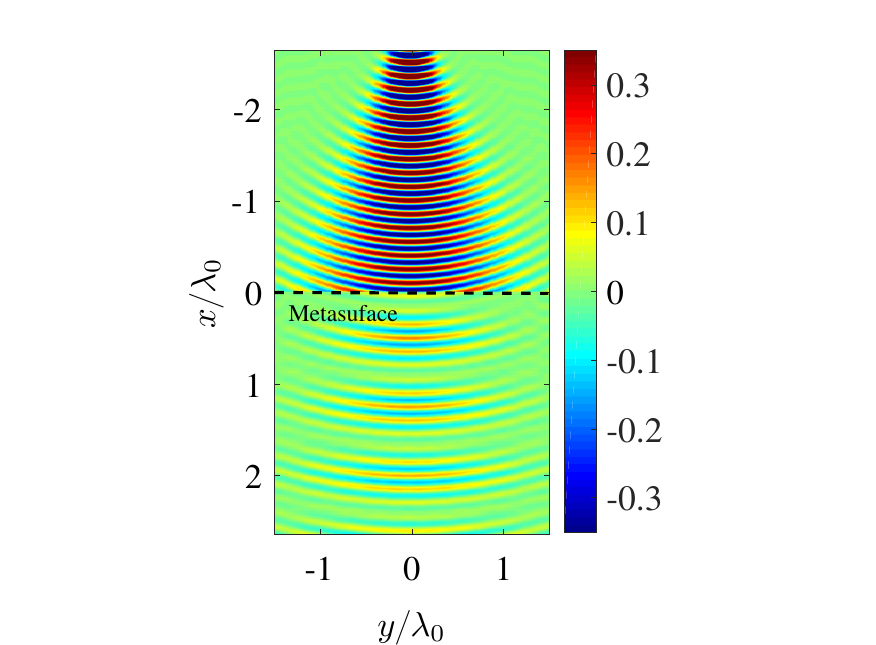}
   \label{Ex5_1_B}}%
  \caption{Example 4: GSTC-FDTD results for the space-time varying metasurface in Fig.~\ref{fig:st_profile}. (a)~Incident and transmitted waves right before ($x=0^-$) and after ($x=0^+$) the metasurface, respectively, at $t=1,823\Delta t$. (b)~Wave pattern in the longitudinal plane of the system with the metasurface positioned at $x=0$ (dashed line). $N_\textrm{res}=200$; the resolution is set so high to account for very small variations in space and time.}\label{FIG:FDTD2DAbsorber}
\end{figure}

The simulation results are shown in Fig.~\ref{FIG:FDTD2DAbsorber}. The incident wave is a normally incident plane wave with gaussian amplitude profile, specifically $E_z(t,y) = \sin\left(2\pi ft\right)\exp(-y^2/0.02)$ with $f=2\:\textrm{GHz}$. The simulation has been run for 30,000 time steps and no sign of instability has been observed. The transmitted and incident fields right after and before the metasurface, respectively, are shown in Fig.~\ref{Ex5_1_A}. The total field is symmetric with respect to the $y=0$ axis since both the metasurface and the incident field in symmetric. The transmitted field amplitude progressively decreases towards the edges of the metasurface, as expected from its specified spatial profile [Fig.~\ref{fig:st_profile}]. The total fields in the transmitted ($x>0$) and incident ($x<0$) regions are plotted in Fig.~\ref{Ex5_1_B}. The time variation of the transmitted field may visualized by inspecting the propagation in the plane of maximal intensity, $y=0$; it fully corresponds to the specifications [Fig.~\ref{fig:st_profile}].

\begin{figure}[ht]
\subfigure[]{%
\includegraphics[width=1\columnwidth]{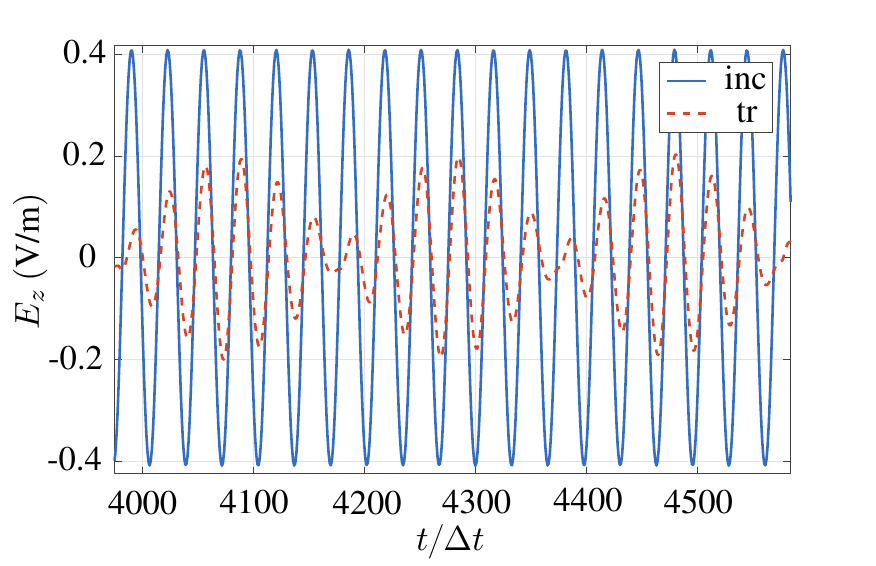}
\label{Ex5_2_A}}%

   \subfigure[]{%
   \includegraphics[width=1\columnwidth]{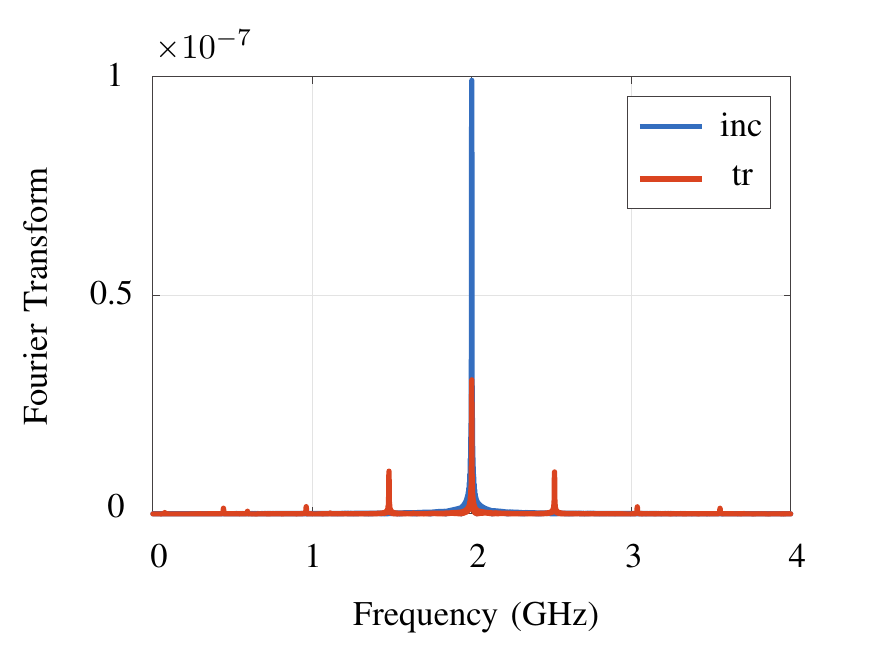}
        \label{Ex5_2_B}}%
  \caption{Example 4: GSTC-FDTD transmitted and incident fields for the space-time varying metasurface of Fig.~\ref{FIG:FDTD2DAbsorber}. (a)~Time variation of the incident and transmitted fields right before ($x=0^-$) and after ($x=0^+$) the metasurface, respectively. (b)~Fourier transform of waveforms in~(a).}\label{FIG:FT}
\end{figure}

Figure~\ref{FIG:FDTD2DAbsorber} plots the temporal [Fig.~\ref{Ex5_2_A}] and spectral [Fig.~\ref{Ex5_2_B}] profiles of the incident and transmitted fields just before and after the metasurface, respectively. Figure~\ref{Ex5_2_A} shows that the transmitted field follows the metasurface periodicity ($200\Delta t$). Figure~\ref{Ex5_2_B} shows the harmonics generated in the transmitted field as a result of time variation, as expected from Sec.~\ref{sec:time_dom}.
\subsection{Example 5}

Finally, we address the 1D problem of the reflection-less half-absorbing and half-transmitting dispersive 0D metasurface. From the previous example, the susceptibilities are $\tilde{\chi}_\textrm{mm}^{yy}=\tilde{\chi}_\textrm{ee}^{zz}=j\frac{2c_0}{3\omega}$. Substituting these susceptibilities into~\eqref{GSTC} yields
 \begin{subequations}
   \begin{align}
     \Delta H_y=\frac{-2c_0\epsilon_0}{3}E_{z,\textrm{av}},\\
     \Delta E_z=\frac{-2c_0\mu_0}{3}H_{y,\textrm{av}}.
   \end{align}
 \end{subequations}
Following again the procedure outlined in Sec.~\ref{sec:analysis_2D} and Sec.~\ref{Sec:Disp_MTs}, the update equation for $H_y^{n+\frac{1}{2}}\left( i_d\right)$ is found as
\begin{align}
   &H_y^{n+\frac{1}{2}}\left(i_d\right)\left(1+\frac{\Delta t c_0}{6\Delta x}\right)=H_y^{n-\frac{1}{2}}\left(i_d\right)\left(1- \frac{\Delta t c_0}{6\Delta x}\right)\\\notag &+\frac{\Delta t}{\mu_0\Delta x}\left[E_z^n\left(i_d+1\right)-E_z^n\left(i_d\right) \right]\\\notag
   &-\frac{\Delta tc_0}{2\Delta x} \left[ H_y^{n+\frac{1}{2}}\left( i_d+1\right)+H_y^{n-\frac{1}{2}}\left( i_d+1\right)\right],
 \end{align}
while that for $E_z^{n+1}\left(i_d+1 \right)$ is
 \begin{align}
   E_z^{n+1}&\left(i_d+1\right)\left(1+\frac{\Delta tc_0}{6\Delta x}\right)=E_z^{n}\left(i_d+1\right)\left(1- \frac{\Delta tc_0}{6\Delta x}\right)+\\\notag &\frac{\Delta t}{\epsilon_0\Delta x}\left[H_y^{n+\frac{1}{2}}\left(i_d+1\right)-H_y^{n+\frac{1}{2}}\left(i_d\right) \right]-\\\notag
   &\frac{\Delta tc_0}{6\Delta x} \left[ E_z^{n}\left( i_d\right)+E_z^{n+1}\left( i_d\right)\right].
 \end{align}

\begin{figure}[ht]
\subfigure[]{%
\includegraphics[width=1\columnwidth]{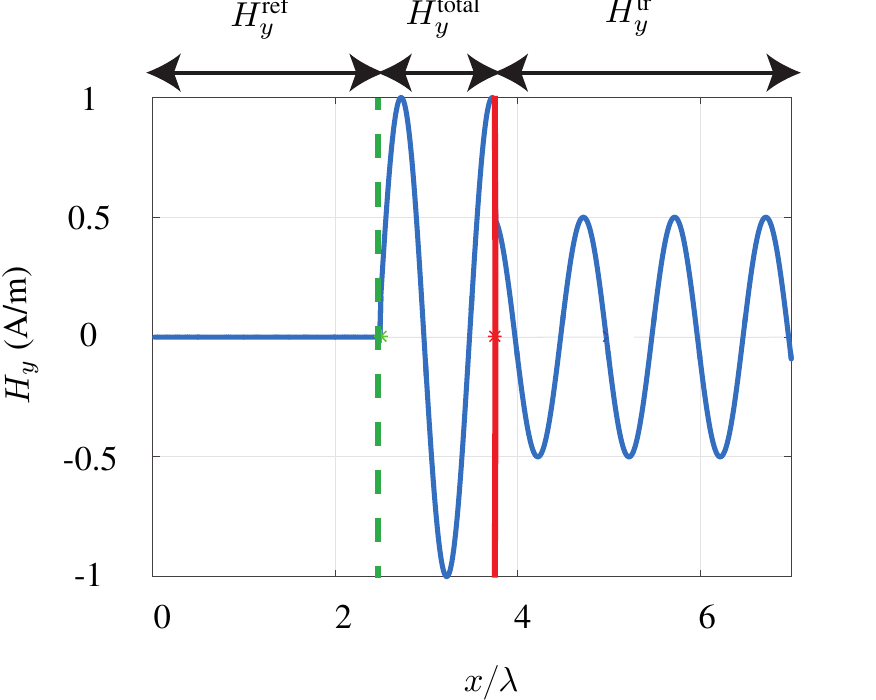}
\label{Ex2_2_A}}%

   \subfigure[]{%
   \includegraphics[width=1\columnwidth]{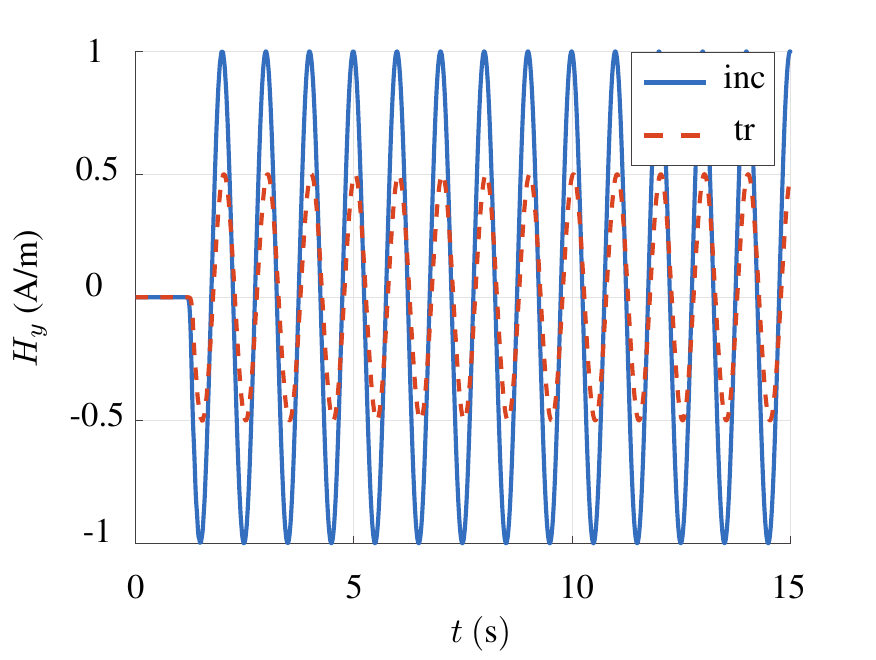}
        \label{Ex2_2_B}}%
  \caption{Example 5: GSTC-FDTD results for the reflection-less half-absorbing and half-transmitting dispersive metasurface. (a)~Spatial profile of the magnetic field waveform at time step 3,000. (b)~Temporal profile of the incident wave right before the metasurface ($x=0^-$) and transmitted wave right after ($x=0^+$) the metasurface. $N_\textrm{res}=30$.}\label{FIG:REFTR}
\end{figure}

Results are shown in Fig.~\ref{FIG:REFTR}. As specified, half of the wave is transmitted and the other half is absorbed by the metasurface. Moreover, the zero reflection expected from the equality of the electric and magnetic susceptibilities is also verified.
\section{Conclusion}\label{sec:conclusions}
We have proposed an FDTD scheme based on Generalized Sheet Transition Conditions (GSTCs) for the simulation of polychromatic, nonlinear and space-time varying metasurfaces. It has been validated and illustrated by five examples. This scheme is fully numerical and very easy to implement, and represents a fundamental extension of the standard FDTD algorithm to general sheet discontinuities. It is currently being extended by the authors to accommodate general metasurface dispersion.

\bibliographystyle{IEEEtran}
\bibliography{GSTCFDTDbib}

\end{document}